\documentclass{eptcs}
\providecommand{\event}{ARCADE 2019}
\usepackage{underscore} 
\usepackage{amssymb}
\usepackage{amsmath}

\title{The Challenge of Unifying Semantic and Syntactic Inference Restrictions}

\author{Christoph Weidenbach
 \institute{Max Planck Institute for Informatics,
  Saarland Informatics Campus E1 4\\
  66123 Saarbr\"ucken, Germany\\
  \email{weidenb@mpi-inf.mpg.de}}}

\def\authorrunning{Weidenbach}

\def\titlerunning{Unifying Semantic and Syntactic Inference Restricions}

\begin{document}

\maketitle

\begin{abstract}
  While syntactic inference restrictions don't play an important role for
  SAT, they are an essential reasoning technique for more expressive logics, such
  as first-order logic, or fragments thereof. In particular, they can result in short
  proofs or model representations. On the other hand, semantically guided
  inference systems enjoy important properties, such as the generation of solely
  non-redundant clauses. I discuss to what extend the two paradigms may be unifiable.
  \end{abstract}



%
%
  
  \renewcommand{\arraystretch}{1.2}
  

  \section{Introduction}

  In~\cite{Weidenbach17} I discussed the differences between simple-syntactic portfolio solvers and
  sophisticated-sematic portfolio solvers. In this paper I present the challenges in combining a
  solver based on syntactic inference restrictions with a solver based on the semantic guidance of inferences.
  More concretely, I investigate the differences between CDCL-style solvers building an explicit partial model assumption
  and solvers based on ordering and selection restrictions on inferences.

\section{NP-Complete Problems}
\label{sect:npcompleteproblems}

The prime calculus for SAT is CDCL
(Conflict Driven Clause Learning)~\cite{MSS96,MoskewiczMadiganZhaoZhangMalik01,BiereEtAl09handbook,BlanchetteEtAl16}.
The calculus can be viewed as a resolution variant where resolution inferences are selected via explicit partial model assumptions.
The CDCL calculus operates on a five tuple $(M,N,U,k,C)$ where $M$ is a sequence of literals representing the current
model assumption, called the \emph{trail}, $N$ the clause set under consideration, $U$ a set of learned clauses, i.e., clauses derived via
resolution, $k$ the number guessed/decided literals in $M$, called the \emph{decision level}, 
and $C$ a clause that is $\top$ if no conflict has occurred yet,
some non-empty clause representing a found conflict, or $\bot$ in case of an overall contradiction.
Consider the clause set

\[ N = \{ P\lor Q\lor R, \; \neg R \lor S, \; \neg S \lor P \lor Q, \; \ldots\} \]

where a CDCL run starting from the empty trail $\epsilon$ may result in

\medskip
$
  \begin{array}{l}
    (\epsilon; N; \emptyset; 0; \top)\\
    \Rightarrow_{\text{CDCL}}^* ([\neg P^1\, \neg Q^2\, R^{P\lor Q\lor R}\, S^{\neg R \lor S}]; N; \emptyset; 2; \neg S \lor P \lor Q)\\
  \end{array}
$

\medskip
where the literals $\neg P$, $\neg Q$ were decided (guessed) at decision levels $1$, $2$, respectively;
$R$ is a propagated literal via the clause $P\lor Q\lor R$; $S$ is propagated via $\neg R \lor S$ and finally in the partial
assignment $[\neg P^1\, \neg Q^2\, R^{P\lor Q\lor R}\, S^{\neg R \lor S}]$ the clause $\neg S \lor P \lor Q$ is false.
The conflict is solved by resolving the false clause with clauses propagating literals from the trail. First with the clause $\neg R \lor S$
resulting in the clause $\neg R \lor P \lor Q$ and then with the clause $P\lor Q\lor R$ resulting in
the clause $P\lor Q$. Now this clause is learned
yielding the new CDCL state

\medskip
$
  \begin{array}{l}
    \Rightarrow_{\text{CDCL}}^* ([\neg P^1\, Q^{P\lor Q}]; N; \{P\lor Q\}; 1; \top).\\
  \end{array}
$

\medskip
Most importantly, the new clause $P\lor Q$ is \emph{non-redundant}, i.e., it is not implied by smaller clauses from $N$, where
the ordering is a lifting of the literal ordering generated by the trail~\cite{AlagiWeidenbach15,FioriWeidenbach19}. Propositional non-redundancy
is itself an NP-complete property. Hence, the CDCL polynomial time model generation procedure either finds a model or eventually
leads to learned clause that enjoys an NP-complete property.
This has several immediate consequences: (i)~termination (ii)~the approach of forgetting clauses works. Note that since only non-redundant clauses
are learned by CDCL, a forgotten clause that has become redundant will not be generated a second time. Hence, learning plus forgetting
can be seen as an efficient way to getting rid of redundant clauses.

Please recall, the classical first-order notion of redundancy means ``not needed to find a proof or model''. Therefore, a redundant
clause can be deleted. In the superposition or ordered resolution context this abstract definition is instantiated with
``not implied by smaller clauses''~\cite{BachmairGanzinger01handbook,NieuwenhuisRubio01handbook}.
However, in the context of SAT, redundancy is often defined as a synonym for
``satisfiability preserving''~\cite{HeuleEtAl19}.
So in the context of many SAT related papers redundant clauses must not be removed, in general, or completeness is lost.
These two notions of redundancy must not be confused and I stick here to the classical first-order notion.

The prerequisites for learning non-redundant clauses are twofold:

\begin{enumerate}
\item Propagation needs to be exhaustive. \label{propexh}
\item Conflict detection needs to be eager. \label{confprio}
\end{enumerate}

The number of literals of a SAT problem is typically small compared to the respective clause set size.
Therefore, exhaustive propagation in SAT is not an efficiency issue, as well as the detection of conflicts, i.e., false clauses.
Surprisingly, what holds for SAT, does not hold for NP-complete problems, in general. It seems that the language
of propositional logic is a nice compromise between expressivity, succinctness, and the efficiency of propagation.

\bigskip
Consider the NP-complete problem of testing satisfiability of a system of linear arithmetic inequations over
the integers~\cite{Karp72,Papadimitriou81}. The language of linear integer arithmetic is more succinct than
propositional logic, i.e., the encoding of on integer variable $x$ requires linearly many propositional variables,
because the a priori bounds for solvability are simply exponential. Now consider a CDCL style procedure
testing solvability of a LIA (Linear Integer Arithmetic) system of inequations~\cite{BrombergerEtAl15}. Consider the example system

\[ N = \{ 1 - x - y \leq 0, \; x - y \leq 0, \ldots\} \]

and a CDCL-style run via CATSAT$+$~\cite{BrombergerEtAl15}, where the partial model assumption is represented by simple bounds, i.e., inequations of the form $x \# c$, $c\in\mathbb{Z}$, $\#\in\{<,>,\leq,\geq\}$.

\medskip
$
  \begin{array}{l}
    (\epsilon; N; \emptyset; 0; \top)\\
    \Rightarrow_{\text{CATSAT$+$}}^* ([x\geq 5^1\, y\geq 6^{1 - x - y \leq 0}\, x\geq 6^{x - y \leq 0}\, \ldots]; N; \emptyset; 1; \top)\\
  \end{array}
$

\medskip
Obviously, the propagation does not terminate, except for a priori simply exponential bounds: if $m$ is the number of inequations
in $N$, $n$ the number of different variables in $N$ and $a$ the maximal coefficient in $N$, then the problem has a solution iff
$-n\cdot (m\cdot a)^{2m+1}\leq x\leq n\cdot (m\cdot a)^{2m+1}$ for every variable $x$ in $N$~\cite{Papadimitriou81}. Hence, there
is currently no CDCL-style procedure known that does exhaustive propagation. Therefore, the learned clauses~\cite{BrombergerEtAl15}
are not non-redundant.
The current state-of-the-art procedures don't use a CDCL-style approach but apply a relaxation of LIA to LRA (linear rational arithmetic) and use LRA solutions
for a branch-and-bound approach. This general idea is complemented via simplifications and fast, sufficient tests for the existence of a
solution~\cite{BradleyManna2007,KroeningStrichmann08,BrombergerWeidenbach17,Bromberger18}.

\section{NEXPTIME-Complete Problems}
\label{sect:nexptimecompleteproblems}

The same phenomenon that occurs at LIA, see Section~\ref{sect:npcompleteproblems}, also shows up if 
the complexity is ``slightly'' increased from SAT to the satisfiability of the Bernays-Schoenfinkel (BS) fragment of first-order logic~\cite{BernaysSchoenfinkel28}, which
is NEXPTIME-Complete~\cite{Plaisted84}. For example, consider the following clause set~\cite{PerezVoronkov08}

\[ N = \left\{\begin{array}{r@{\; : \;}l}
    1 & P(0,0,0,0),\\
    2 & \lnot P(x_1,x_2,x_3,0) \lor P(x_1,x_2,x_3,1),\\
    3 & \lnot P(x_1,x_2,0,1) \lor P(x_1,x_2,1,0),\\
    4 & \lnot P(x_1,0,1,1) \lor P(x_1,1,0,0),\\
    5 & \lnot P(0,1,1,1) \lor P(1,0,0,0),\\
         6 & \lnot P(1,1,1,1)\\
       \end{array}\right\}
 \]

 where a CDCL-style run, using ground literals for the partial model assumptions~\cite{FioriWeidenbach19}, will propagate all values of the 4-bit counter represented by $P$:

 \medskip
$
  \begin{array}{l}
    (\epsilon; N; \emptyset; 0; \top)\\
    \Rightarrow_{\text{SCL}}^* ([P(0,0,0,0)^{C1}\, P(0,0,0,1)^{C2}\, P(0,0,1,0)^{C3}\,\ldots P(1,1,1,1)^{C2}],N,\emptyset,0,\lnot P(1,1,1,1))\\
  \end{array}
$

\medskip
\noindent
before detecting the conflict, where for the propagation justifications the notation $C<\text{clause number}>$ was used.
Thus, there are exponentially many
propagations, in general. Still, the clause set can be refuted in linearly many steps by starting with resolution steps between the clauses $2-4$:

\medskip
$\begin{array}{lr@{\; : \;}l}
    2.2 \;\text{Res}\; 3.1 & 7 & \lnot P(x_1, x_2, 0, 0) \lor P(x_1,x_2,1,0)\\
    7.2 \;\text{Res}\; 2.1 & 8 & \lnot P(x_1, x_2, 0, 0) \lor P(x_1,x_2,1,1)\\
    8.2 \;\text{Res}\; 4.1 & 9 & \lnot P(x_1, 0, 0, 0) \lor P(x_1,1,0,0)\\
    9.2 \;\text{Res}\; 8.1 & 10 & \lnot P(x_1, 0, 0, 0) \lor P(x_1, 1, 1, 1)\\
    10.2 \;\text{Res}\; 5.1 & 11 & \lnot P(0, 0, 0, 0) \lor P(1,0,0,0)\\
    11.2 \;\text{Res}\; 10.1 & 12 &  \lnot P(0, 0, 0, 0) \lor P(1, 1, 1, 1)\\
    12.1 \;\text{Res}\; 6.1 & 13 & \bot.\\
 \end{array}$

 where this proof can be implemented by a standard ordering restriction on the $P$ atoms, e.g., via a KBO (Knuth Bendix Ordering),
 plus a respective
 selection strategy~\cite{BachmairGanzinger94b}. For example, for the first step the second literal out
 of clause $2$ is maximal with respect to a KBO instance where $1 >_{\text{KBO}} 0$ and the first literal of clause $3$ is selected.

 The above clause set $N$ without clause~6 is obviously satisfiable. Still, a CDCL-style calculus using ground literals generates
 exponentially many ground atoms before it detects satisfiability. An ordered resolution calculus using a standard KBO
 instance where $1 >_{\text{KBO}} 0$ does not generate any clause at all, because exactly the positive literals are maximal.
 Extending the model representation language from ground atoms~\cite{FioriWeidenbach19} to more a more expressive language
 including variables~\cite{BaumgartnerFuchsTinelli06,FermuellerPilcher07,PiskacEtAl10,AlagiWeidenbach15,BonacinaPlaisted16}
 solves the issue with the above example by starting with a model assumption $[P(x_1,x_2,x_3,x_4)^1]$. However, the more
 expressive model language requires more complex operations in order to guarantee consistency of the model assumption and
 to detect propagating literals and false clauses. The discrepancy between a compact model representation and complex calculations
 cannot be resolved in general, because $\text{NP} \neq \text{NEXPTIME}$.

 In addition, for ``practically relevant'' instances of the BS fragment the situation is the same. We have shown~\cite{SudaEtAl10}
 that the YAGO~\cite{SuchanekKW07} fragment of BS can be effectively saturated by a variant of ordered resolution with chaining,
 whereas all model-guided approaches fail. Again for the reason that propagation with respect to millions of constants cannot be
 efficiently done, yet.

 \section{Unification}

 The two frameworks, syntactic inference restrictions and model-guided inferences cannot be easily combined.
 Obviously, since ordered resolution may generate redundant clauses, but model-guided inferences with exhaustive
 propagation and eager conflict detection do not, model-guided inferences cannot simulate resolution inferences.
 The resolution inferences in Section~\ref{sect:nexptimecompleteproblems} refuting the clause set encoding a counter are not redundant.
 Still, they cannot be simulated by a CDCL-style calculus~\cite{FioriWeidenbach19} because it will immediately run
 into the exponentially many propagation steps and find a refutation of exponential size this way. It seems to
 be non-trivial and non-obvious how the two paradigms can be unified. If eager propagation is dropped, then
 model-guided inferences can simulate resolution~\cite{PipatsrisawatDarwiche09,FioriWeidenbach19}, however, in this case
 non-redundancy of learned clauses is lost, in general.

 One way out of this dilemma could be to limit the amount of propagations
 by limiting the number of literals that may be derived by propagation. For example, the InstGen calculus~\cite{GanzingerKorovinEtAl03}
 limits ground instantiation and the generation of new ground literals to using exactly one constant. Then the potential size of a model
 assumption remains linear in the size of the investigated clause set. However, a terminating model assumption search does not
 result in an overall model for the clause set anymore. Thus the model generation process itself needs to turn into a ``learning'' procedure.

\bibliographystyle{eptcs}
\bibliography{paper}

\end{document}